\documentclass[12pt]{iopart}

\usepackage{xspace}
\usepackage{epsfig}
\usepackage{cancel}

\usepackage{graphicx}

\newcommand{\nc}{\newcommand}

\nc{\beq}{\begin{equation}}
\nc{\eeq}{\end{equation}}
\nc{\beqa}{\begin{eqnarray}}
\nc{\eeqa}{\end{eqnarray}}
\nc{\bea}{\begin{eqnarray}}
\nc{\eea}{\end{eqnarray}}
\nc{\ra}{\rightarrow}

\newcommand{\dalam}{\raise-1mm\hbox{\large$\Box$}}

\nc{\rhob}{\rho_{\rm b}}
\nc{\rbdm}{R_{\rm b/dm}}
\nc{\rhodm}{\rho_{\rm dm}}

\def\lsim{\mathrel{\raise.3ex\hbox{$<$\kern-.75em\lower1ex\hbox{$\sim$}}}}
\def\gsim{\mathrel{\raise.3ex\hbox{$>$\kern-.75em\lower1ex\hbox{$\sim$}}}}
\newcommand{ \slashchar }[1]{\setbox0=\hbox{$#1$}   
   \dimen0=\wd0                                     
   \setbox1=\hbox{/} \dimen1=\wd1                   
   \ifdim\dimen0>\dimen1                            
      \rlap{\hbox to \dimen0{\hfil/\hfil}}          
      #1                                            
   \else                                            
      \rlap{\hbox to \dimen1{\hfil$#1$\hfil}}       
      /                                             
   \fi}                                             %

\begin{document}

\title{On Relating the Genesis of Cosmic Baryons and Dark Matter}

\author{Hooman Davoudiasl}
\ead{hooman@bnl.gov}
\address{Department of Physics,
Brookhaven National Laboratory, Upton, NY 11973-5000, USA}

\author{Rabindra N. Mohapatra}
\ead{rmohapat@umd.edu}
\address{Maryland Center for Fundamental Physics, Department of Physics,
University of Maryland, College Park, MD 20742, USA}

\begin{abstract}

The similar cosmological energy budgets in visible baryons and dark
matter motivate one to consider a common origin for the generation
of both. We outline the key features of scenarios that
can accommodate a unified framework for the genesis of cosmic
matter. In doing so, we provide a brief overview of some of the past
and recent developments and discuss the main predictions of a number of 
models.

\end{abstract}

\maketitle


\section{Introduction}
\label{intro}

The nature of matter has been a question of fundamental import in
science and philosophy, for centuries.  While the initial inquiries
of antiquity had more of a philosophical character, it was the
application of the scientific method in probing Nature that brought
us a firm understanding of matter.  Over the last century,
experimental examination of the structure of matter at
ever-decreasing length scales culminated in the emergence of the
Standard Model (SM) of particle physics.

The SM provides a microscopic description of the visible matter in
the world around us. However, in parallel, over the last several
decades, mounting evidence from various astronomical observations
have led us to reach a surprising conclusion: The visible matter,
most of whose mass is composed of baryons, is in fact responsible
for about $5\%$ of the cosmic energy density, while the dominant
material mass in the Universe, comprising about $22\%$ of its energy
budget \cite{Komatsu:2010fb}, is ``dark" and does not have any appreciable interactions
with the visible matter. Hence, while the SM is our most precise
theory of Nature, it only describes a small fraction of what makes
up the cosmos!  In fact, the situation is worse: even the
visible content of the Universe, made up of baryons and almost devoid
of anti-baryons, requires a baryogenesis mechanism to generate the
requisite baryon asymmetry and it is widely believed that successful
baryogenesis requires extending the SM.

Thus, it seems that our latest understanding of cosmology has left
us with two unresolved puzzles: (1) the nature of dark matter (DM)
and (2) the origin of the baryon asymmetry in the Universe. The
answer to the first question in some of the most popular scenarios
of physics beyond SM is that DM is made up of a stable particle
whose relic density is set by thermal freeze-out \cite{Feng:2010gw}.  That is, as the
Universe cooled down after the Big Bang, the interactions that
annihilated DM particles got less and less efficient and at some
point decoupled, leaving a relic DM population.
It turns out that weak scale interactions, characterized by masses of
order 1~TeV, roughly give the correct order of
magnitude for the DM relic density.  Given
the importance of the weak scale in particle physics, neutral and
stable Weakly Interacting Massive Particles (WIMPs) are popular
candidates for DM and have been the focus of much theoretical, as
well as experimental, activity.  The answer to question (2) above,
however, requires the introduction of a mechanism that results in a
baryon asymmetry and is apparently unrelated to the physics that sets
the relic density of WIMPs.  Hence, it seems that the visible and
dark material contents of the cosmos are set by disjoint mechanisms.

The above discussion raises an intriguing question: why would then
two seemingly unrelated sectors end up having similar contributions
to the energy density of the Universe?  This question, which is
based on firm empirical evidence, leads us to examine whether
baryons and DM could have a common origin.  In particular, since the
relic density of baryons is set by an asymmetry, one may naturally
conclude that the DM cosmic abundance was similarly 
obtained \cite{Hut:1979xw,Nussinov:1985xr,BCF}.
In recent literature, this has been called the 
Asymmetric Dark matter (ADM) hypothesis and we will 
use this terminology throughout this article.  
Typical implementations of the ADM hypothesis
generally yield similar number densities for the visible and dark
matter populations and we give some examples below. 
It is clear that these theories may quite naturally be characterized by
DM particles whose masses are not much larger than that of the
proton, $m_p\sim 1$~GeV. The properties of ADM can then be quite
distinct from WIMPs and motivate different search strategies, as we
will discuss later.

In the next section we will outline some of the key features of
typical ADM scenarios that provide a unified framework for the
generation of both baryonic and dark cosmic matter. In section
\ref{props}, we will discuss some of the early and recent proposals,
representing different classes of ADM models.  In section
\ref{pheno}, we briefly consider the phenomenological aspects of
different classes of ADM models, and the search strategies that they
motivate.  We discus some of the astrophysical implications of ADM scenarios in section \ref{astro}.  
A summary and some concluding remarks are provided in
section \ref{summary}.

Before closing this introduction, we would like to add that this
article, given its length and scope, is not meant to be a
comprehensive review of the literature.  As a result, many
interesting ideas, directly or indirectly relevant to the topic,
could not be covered by our review; their omission is not an
indication of their lesser significance.  However, it is our hope
that this brief survey of the subject can be a helpful reference for
some of the key ideas and questions associated with unified theories of
cosmic baryons and dark matter.

\section{The Main Ideas}
\label{Ideas}

As already mentioned, the similar energy budgets in baryons and DM
motivates a unified theory for their origin. While the ratio of
baryon energy density $\rhob$ to the critical energy density
$\rho_{\rm c}$ is $\Omega_{\rm b} \simeq 0.05$, the same ratio for
the dark matter energy density $\rhodm$ is $\Omega_{\rm dm} \simeq
0.22$ \cite{Komatsu:2010fb}.  For the purpose of discussions that follow, let us define
\beq
\rbdm = \frac{\Omega_{\rm b}}{\Omega_{\rm dm}},
\label{rbdm}
\eeq
where $\rbdm \simeq 0.2$.  Observational data, as well as theoretical arguments, strongly
indicate that the visible matter in the Universe is dominated by
baryons and that the cosmic anti-baryon density
is negligible in comparison \cite{KT}.  Any
mechanism for the generation of baryon asymmetry of the Universe
(BAU) needs to have ingredients that allow for the implementation of
the three Sakharov conditions \cite{Sakharov:1967dj}: 
(1) baryon number violation, (2) C and 
CP violation, and (3) departure from equilibrium.  Condition (1)
is an obvious requirement, while condition (2) ensures the
underlying physics can distinguish between matter and anti-matter.
The last condition is needed to avoid washing out the asymmetry
generated by the first two, when various processes and their
inverses are in equilibrium in the early Universe.

A unified mechanism for generation of visible and dark matter must
then accommodate the above criteria.  This can be arranged in a
variety of ways.  However, many models fall in one of the following
two main categories:

\indent{(I) Models where a quantum number is assigned to DM that is
also shared by the visible matter.  An asymmetry in this number must
then be shared between the two sectors through certain
interactions. Once these interactions decouple from the thermal
plasma, asymmetries of comparable size get frozen in the visible and
dark sectors.} \vskip0.25cm

\indent{(II) Models in which the concept of baryon number $B$ is
generalized, with equal and opposite asymmetries sequestered in the
visible and DM particles. This can happen, for example, if the theory
including the dark particles has a symmetry generated by a charge
$Q_{tot}= B+Q_X$ , where $Q_X$ is the charge associated with the $U(1)$ 
symmetry in the dark matter Lagrangian.  In this class of models, 
while $B+Q_X$ is always preserved, the orthogonal combination
$B-Q_X$ is broken and is responsible for the asymmetry generation. 
The latter guarantees that $n_B=-n_{Q_X}$.
Once the asymmetries are produced out-of-equilibrium,
processes that can wash them out should remain decoupled 
and relic baryon and DM densities persist.  However, the net ``baryon
number" of the Universe remains zero in these scenarios.}

In either class of ideas, one needs to ensure that processes that
annihilate the symmetric population of particles and their
anti-particles are efficient, so that the relic densities are set
only by the asymmetries.  Note that visible baryons have strong
interactions that easily accomplish this, while generic DM sectors
are not guaranteed to have the requisite interactions. Furthermore,
ADM can be both of bosonic or fermionic type.

While the experimental signatures of ADM models of types (I) and (II) could cover a 
wide range of possibilities, under some general assumptions, certain characteristic 
features may be ascribed to each type of model.  For instance, in Ref.~\cite{Ibe:2011hq}, 
assuming that the shared quantum number in type (I) models is $B-L$, a generic 
relation $m_{\rm dm} \sim (5-7){\rm ~GeV}/q_{\rm dm}$ 
between the charge $q_{\rm dm}$ and the mass $m_{\rm dm}$ of ADM 
is obtained, while in Ref.~\cite{vonHarling:2012yn}, the typical 
relation $m_{\rm dm} \sim q_{\rm dm} (2-5)$~GeV is derived for models of type (II) (with $B\to B-L$ in the above).  
We see that under the general assumptions in Refs.~\cite{Ibe:2011hq,vonHarling:2012yn}, and also assuming 
$q_{\rm dm} \lsim 1$, one could expect models of type (I) to yield ADM masses that are typically larger 
than those associated with type (II) models.  We note that the expected DM mass range 
is an important input for choosing a search strategy.  For example, direct detection of dark matter based on 
measuring nuclear recoil in  DM-nucleus scattering  
becomes less efficient for low values of $m_{\rm dm}$ and alternative approaches may need to be devised 
if one expects $m_{\rm dm}\lsim 1$~GeV \cite{Essig:2011nj}.   

Finally, we note that a complete explanation of the similarity of dark and visible matter energy densities
will also require an understanding of why the mass of the dark matter particle is similar to that 
of nucleons. Most models do not address this issue except for the mirror model discussed below.

\section{Some Early and Recent Proposals}
\label{props}

\subsection{Class (I) models}

\subsubsection{Technibaryonic ADM}

The notion of a stable baryon can easily be accommodated in
composite models, such as technicolor, where fermions can be bound
into analogues of protons by weak scale strong dynamics. Thus both sectors,
the technicolor and SM, can share a common baryon number and if techni-baryons 
are dark matter, these models will fall into 
class I models using our classification above.  Indeed, the
earliest proposals for asymmetric dark matter  \cite{Nussinov:1985xr,BCF} were based
on technicolor models.  For example, in Ref.~{\cite{BCF} it was
proposed that fermion number violating interactions in the early
Universe \cite{Kuzmin:1985mm}, often referred to as sphalerons
\cite{Arnold:1987mh}, can distribute asymmetries over the entire
electroweak sector, including the techni-fermions.  In this case,
a similar asymmetric number density of quarks and techni-quarks can
be produced, as can be seen by solving for the relevant chemical
potentials and imposing neutrality conditions \cite{HT}. If the
lightest techni-baryon is neutral under the SM interactions then it
could be a suitable DM candidate as long as it is stable on
cosmological time-scales.  This scenario is of the type in category
(I) above, as noted. Extensions of this idea in other theories of strong
dynamics at the weak scale that address precision 
electroweak data \cite{NewTC} and may 
give rise to potential astrophysical signals \cite{Nardi:2008ix}
have been proposed in recent years.  At first, it may
seem that theories based on technicolor would lead to an
unacceptable DM energy density, since techni-baryons in these models
are expected to have masses $m_{TB}\sim 1$~TeV.  Obviously, if
baryons and DM develop similar densities in such models, one would
end up with an energy density in DM much larger than implied by the
data. However, this issue can be addressed through the same
fermion-number violating sphaleron processes that lead to the
asymmetries.  To see this, note that the temperature at which the
sphalerons decouple is typically of order the electroweak phase
transition temperature $T_c \sim 100$~GeV.  If the techni-fermion changing 
processes stay in equilibrium below the techni-baryon mass, we
generically expect a suppression in techni-baryon number of order
$(m_{TB}/T_c)^{3/2} e^{-m_{TB}/T_c}$ \cite{BCF,NewTC}.  For typical
values of $m_{TB}$ and $T_c$, one can then get an ADM number density
suppression of order $10^{-3}$ to $10^{-2}$, and end up with an
acceptable DM energy density.

\subsubsection{Models based on $B$ or $B-L$ charge}

A second class of proposals in category (I) are not based on the
assumption of electroweak symmetry breaking (EWSB)
via strong dynamics.  An early example is
Ref.~\cite{Kaplan:1991ah} that uses extra electroweak fermions
charged under an anomalous $U(1)_X$ global symmetry.  In this model,
it is assumed that EWSB occurs through a first order phase
transition and that the new fermions have CP violating interactions
with the bubble wall separating the symmetric and broken phases in
the plasma.  As a result, a net charge is transported into the
symmetric phase that electroweak sphalerons process into baryon and
DM asymmetries.  The DM candidate here is the lightest particle
charged under the $U(1)_X$, whose mass is near the weak scale. Here,
the ratio $\rbdm$ is obtained by the ratio of the scales of proton mass and
the weak scale, up to a factor of order unity determined by the anomaly equation.
A main feature of this model is the
use of a quantum number that gets partitioned between the visible
and and the dark matter sectors through the effect of certain
interactions. For example,  a net $B-L$
is assumed to be generated in the model of Ref.~\cite{Kaplan:2009ag},
at a high temperature, but preserved at
lower temperatures, and transferred to a DM sector that carries
$B-L$ charge. If the transfer operators decouple above the mass of
the DM particle, a DM asymmetry of the same order as the baryon
asymmetry is generated and preserved. Such a scenario then predicts
that the DM particle has a mass $5-15$~GeV. Note that a net $B-L$ in
the SM fermions can get processed into non zero $B$ and $L$
asymmetries by sphalerons in thermal equilibrium \cite{HT}.

\subsubsection{Mirror matter models}

Another  example of type (I) models, but with very distinct features, 
is that of Ref.~\cite{An:2009vq} which is based on the idea that there 
may be two parallel sectors in the Universe
with identical matter and force content related by a discrete $Z_2$ symmetry 
(parity) with gravity and other SM singlet
fields connecting the two sectors. These models
are known in the literature as mirror models (for a review and extensive 
references to literature prior to 2007, see Ref.~\cite{Okun:2006eb}). The presence of the discrete mirror symmetry implies that
all couplings in the mirror sector are the same as those of the SM, prior to symmetry breaking.
This is a unique feature of this model since it helps to prevent the proliferation of 
coupling parameters in the theory. In fact prior to 
EWSB in both sectors, the  parameters of the entire model are those of
the SM. Once symmetry is broken, new parameters associated with 
symmetry breaking vacuum expectation values (vevs) 
appear. This lends a certain degree of economy and predictivity to these models.  
The new features that help to connect visible and dark matter in these models
are the following: (i) the two sectors are connected by singlet right handed neutrinos 
$N_a$, $a=1,2,3$, whose couplings are given by:
\begin{eqnarray}
{\cal L}_I~=~h_{\nu, a}\bar{N}_a(LH~+~L'H') + h.c.
\end{eqnarray}
where $L$ and $H$ denote the SM lepton and Higgs doublets, with corresponding
mirror fields, denoted by a prime. This makes 
the lepton number of the two sectors the same, which in turn makes it
a quantum number sharing model of type (I). 
One then adds a Majorana mass for the singlet  neutrino fields  $N_a$  
\cite{An:2009vq} which breaks this common lepton number.
Due to the presence of CP violation in the Yukawa couplings $h_{\nu,a}$, 
when the above Yukawa interactions go out of equilibrium, 
leptogenesis occurs \cite{Fukugita:1986hr} creating a lepton 
asymmetry in both the familiar and the mirror sector. At the tree level, due to mirror symmetry,
the two lepton asymmetries are equal. The symmetry breaking vevs which may be different
in the two sectors do not affect this equality since leptogenesis occurs much above the symmetry breaking temperatures.
Another way to see this equality is to note that  the effective interaction generated after $N_a$ 
decouples, is $LHL'H'$ which conserves the quantum number,
$L-L^\prime$. The lepton asymmetry in both sectors are 
subsequently converted to baryon asymmetry by the SM 
sphalerons and their mirror analogs.
The mirror baryons are dark matter in these models and their abundance 
is equal to the observed baryon asymmetry. 
There may be some small differences between the asymmetries once radiative corrections are
taken into account.  It is important to point out that 
the symmetric part of the dark matter abundance gets 
annihilated by the mirror analog of the SM strong (QCD) 
interactions and no new postulate is needed.

An important characteristic of this class of models is that one can provide
a rational for the dark matter mass being slightly higher (but of the same order of magnitude) than the
familiar baryons. The way to see this is as follows: when the mirror weak scale is
made higher than the visible sector weak scale, running of the mirror coupling change and
if both couplings were grand unified at some high scale, the weak scale asymmetry will imply that
the mirror QCD scale, $\Lambda^\prime$ becomes non-perturbative at a much higher scale than
the $\Lambda_{QCD}$ of the visible sector. This coupled with higher quark masses of the mirror sector
(due to larger $v^\prime_{wk}$), implies that lightest baryon of the mirror sector is a few times larger baryon mass
compared with the visible sector (for details see \cite{An:2009vq}).

Prior to symmetry breaking this model has double the number 
of light particles in the SM. Therefore, one way to make these models  compatible with the constraints
of big bang nucleosynthesis (BBN) is to make the three mirror neutrinos and the mirror photon 
massive so that they can decay before the BBN. This can be achieved by suitably
choosing the symmetry breaking in the mirror sector. 

Furthermore, since  a priori, a kinetic mixing between the
familiar photon $\gamma$ and the mirror photon $\gamma'$ 
is allowed by gauge invariance, this could be included in the Lagrangian, thereby 
connecting the familiar and the mirror sectors at lower temperatures.  
This mixing has the consequence that it can lead to signals in direct detection searches for mirror dark matter. 
The existence of  $\gamma$-$\gamma^\prime$ mixing can be tested at accelerators 
and we will discuss this below.

Other variations of this idea may also exist {\it e.g.} 
one could assume a real scalar field with couplings
\begin{eqnarray}
{\cal L}_I\sim \frac{1}{\Lambda^6} S[(u^cd^cd^c)^2+(u^{\prime,c}d^{\prime,c}d^{\prime,c})^2]
\end{eqnarray}
where we have suppressed the generation index.  
CP violation in the $S$ couplings can then allow this operator to
generate equal baryon asymmetry directly in both sectors 
without the intervention of sphalerons. Mirror baryons, {\it e.g.} the mirror neutrons $N'$, 
make up the dark matter.  An intriguing possibility in this scenario
is that over the real long term future of the universe, dark matter could scatter into the visible sector via  $N'+N'\to N+N$ ``emptying the universe of all its dark matter."

\subsubsection{Other examples} 

There have been many other models for ADM which have the property 
of quantum number sharing between the visible and dark sector as their basis.  
For example, Ref.~\cite{Shelton:2010ta} assumes that DM asymmetry is 
generated by ``dark sphalerons" of a hidden non-Abelian gauge group, 
during a first order phase transition.  If the dark phase transition occurs below the temperature 
for electroweak phase transition, the asymmetry in the dark matter $X$ must be transferred to SM baryon directly.
In a supersymmetric realization, this can be achieved via superpotentials of type 
$(1/\Lambda^2)X^2 u^c d^c d^c$ suppressed by a scale $\Lambda$.  
However, if the dark phase transition takes place before electroweak phase 
transition one may use electroweak sphalerons to achieve the transfer 
of the asymmetry to the visible sector.  The typical mass of the ADM particle in this framework is in the 1-5~GeV range.  For recent work that allow for much heavier ADM particles near the weak scale see, for example,  
the proposals in Ref.~\cite{Buckley:2010ui} where DM number density is thermally suppressed, or
Ref.~\cite{Chun:2010hz}, where a weak Dirac gaugino is the DM in the context of supersymmetric models.  

In this class of ideas, there is another  model \cite{GS} where the DM particle $\chi$
has fractional lepton number so that it is stable. In this model,
some $L=0$ heavy scalar decays into $\chi$ 
as well as other lepton number carrying scalars
such as the SM triplet Higgs boson $\Delta$ responsible for neutrino 
masses via type II seesaw mechanism.  In the presence of CP violation,
these decays generate an asymmetry between $\chi$ and $\bar{\chi}$  
and an asymmetry of the same order between  $\Delta$ and 
$\bar{\Delta}$. The $\chi$ asymmetry stays as the dark matter 
whereas the $\Delta$ asymmetry translates to a lepton number asymmetry
when $\Delta$ particles decay. The lepton number asymmetry 
then gets converted to baryon asymmetry via the electroweak sphaleron interactions. 

There are also proposals that connect the genesis of visible and dark cosmic matter through 
the formation baryon-number-carrying condensates, as in the 
Affleck-Dine baryogenesis mechanism \cite{Affleck:1984fy}.  Such 
a condensate may later evolve into non-topological solitons \cite{Kusenko:1997si} called $Q$-balls, originally   
introduced by Coleman \cite{Coleman:1985ki}, which can arise in supersymmetric 
extensions of the SM \cite{Kusenko:1997zq}.   
When $Q$-balls  have sufficiently large baryonic (or leptonic) charges they can 
be cosmologically long-lived and provide a contribution to dark matter \cite{Kusenko:1997si}. 

For other works on ADM and baryogenesis see Refs.~\cite{KL1,KL2,dutta,others}.  
We wish to point out that there is a whole class of ADM models 
where there is no direct connection between baryogenesis and dark matter genesis \cite{otherADM}.  Also, note that 
in some models \cite{non-ADM} 
the generation of baryons and the abundance of dark matter may be controlled by the same underlying 
considerations, without resulting in an ADM scenario.

\subsection{Class (II) Models}

These models are based on the possibility that the Universe has a net zero baryon number that is 
generalized to encompass both the visible and the dark sector.  An early realization of this idea was 
proposed in Ref.~\cite{Dodelson:1989ii}, using a scalar condensate that stores anti-baryon number 
and can act as cold DM in the current epoch.  Here, one can 
think of cosmic DM content as being effectively ``anti-matter," storing a 
baryon number that is equal and opposite to that sequestered in SM nucleons \cite{Dodelson:1989ii}.  
These basic features are characteristic of class (II) models and have been implemented in a variety 
of other proposals \cite{Kuzmin:1996he,Farrar:2004qy,Agashe:2004bm,Gu:2007cw,gu,hylo,Bell:2011tn}.

For instance, in Ref.~\cite{hylo}, it was proposed that out-of-equilibrium decays of heavy Dirac fermions $X$ 
lead to the simultaneous generation of a baryon asymmetry and a dark matter asymmetry.  This is achieved 
through Yukawa couplings of $X$ to a hidden fermion $Y$ and a scalar $\Phi$, and mass-suppressed couplings to 
``neutron" operators $u^c d^c d^c$, with $u$ and $d$ denoting up- and down-type quarks, respectively.  In this scenario, 
equal and opposite baryon asymmetries are stored in nucleons and a population of $(Y,\Phi)$ particles, whose sum of 
masses is around 5~GeV.  In order to avoid washout effects, the reheat 
temperature is low, around or below $\sim 1$~GeV.  
Here, symmetries and mass relations ensure the stability of both types of baryon, 
but this does not preclude {\it induced nucleon decay} (IND), that is the 
destruction of SM baryons in scattering from the cosmic DM population, leading to 
interesting phenomenological signatures \cite{hylo,IND}; see also Ref.~\cite{Farrar:2004qy} 
for a different model where such effects were considered.  
In this model, IND could be a striking signature in nucleon decay 
experiments, while direct detection based on nucleon recoil experiments may be suppressed, 
depending on the strength of the mechanism for 
DM symmetric annihilation into SM states.

Another example of type (II) models is that of Ref.~\cite{gu}, 
where one introduces color-charged and color singlet but
baryon number carrying particles at the TeV scale. 
The color-charged particles couple to the SM particles as well as to the color singlet particles.
The lightest color singlet particle plays the role of dark matter. The heavy color singlet particle decays to both
SM color-charged states as well as to dark matter. In the presence of CP violation, this will generate both ADM relic density
as well as a baryon asymmetry relating both of them in the process, 
via the new couplings of the exotic states.

We close this section with a couple of comments.  First of all, it was 
noted in Ref.~\cite{Buckley:2011ye} that if a theory of ADM admits
interactions that change dark matter to anti-dark matter, then DM-anti-DM oscillation \cite{Cohen:2009fz} 
will remove all dark matter from the Universe making the model 
not viable for description of dark matter, if the oscillation time is of order or less than the age of
the universe. This point has been reanalyzed in Ref.~\cite{Cirelli:2011ac}, where it is noted that asymmetric dark matter
in the presence of possible DM-anti-DM oscillations remain viable if the dark matter mass is
between 100-1000 GeV.  Also, the analysis in Ref.~\cite{Tulin:2012re} indicates that whether DM-anti-DM oscillations 
result in the resumption of annihilations depends on the type of DM interactions with lighter fields.    
In many dark matter models however ({\it e.g.} mirror DM)
such interactions are forbidden by specific symmetries (such as mirror baryon number in mirror models).

Secondly, our classification of models of ADM into types (I) and (II) is meant to be taken as a general guide.  However, 
there are models that incorporate features of both classes and are not clearly of one type or the other; see for example 
Ref.~\cite{Graesser:2011vj}.


\section{Phenomenology}
\label{pheno}
We now turn to phenomenological implications of ADM 
models and comment on some astrophysical implications as well.  Here again we focus on some generic
consequences rather than model specific ones with the goal of  distinguishing an ADM model from the conventional WIMP hypothesis. We will discuss direct detection searches, signals at colliders, novel probes, and some indirect tests.  

\subsection{Direct Detection and dark photon} 
Direct detection of dark matter requires that there must be particles in the 
theory that interact with both the SM particles as well as the dark matter particle. For example in the 
minimal supersymmetric Standard Model (MSSM), the lightest supersymmetric particle, which is DM, couples to the $Z$ boson whose interactions with quarks lead to a DM signal.  In the ADM models, the situation is somewhat more complex.  
First of all, experiments designed for direct detection of weak scale ($\sim 100$~GeV) WIMPs generally do not have 
high sensitivity to signals from typical ADM particles characterized by GeV scale masses.  In addition, it is possible to have ADM models where dark matter is completely invisible to direct searches.  An example of this kind of model is the mirror ADM model where the familiar photon and the mirror photons do not mix \cite{otherADM}. However, one can supplement these models with the gauge invariant kinetic mixing $\epsilon B^{\mu\nu}B^\prime_{\mu\nu}$ \cite{holdom}, 
between the SM $U(1)_Y$ and mirror $U(1)_{Y'}$ gauge fields.  Such operators arise in other setups 
that include an additional $U(1)'$ gauge interaction, for example as may be  
required for symmetric annihilation in ADM scenarios \cite{hylo}.  In any event, direct 
detection may still be quite suppressed even when $\epsilon$ is sufficiently large for this purpose \cite{hylo}.
The kinetic mixing parameter is however subject to different constraints, 
depending on whether the mirror (or dark) photon is massive or massless.    

\subsubsection{Massive dark photon}
In this class of models, in addition to the kinetic mixing term, 
there is a mass term for the mirror photon, $m_{\gamma'}$, which may arise out of spontaneous
breaking of mirror electromagnetic gauge invariance. 
Gauge invariance associated with familiar electromagnetic 
$U(1)$ of course remains unbroken \cite{An:2009vq}. In this case,
there are constraints from supernova 1987A observations, 
if $m_{\gamma'} \leq 100$ MeV, set by the core temperature ($\sim 30$~MeV) 
of the supernova in the initial stages of the explosion.
The limit \cite{Kolb:1996pa} is $\epsilon \leq 10^{-9.5}$. 
There are other bounds on this from other considerations: {\it e.g.} there are constraints, from measurements of cosmic 
background radiation, of $\epsilon < 10^{-7 }- 10^{-5}$, 
for hidden photon masses between $10^{-14}$  eV and $10^{-7}$ eV \cite{Mirizzi:2009iz}.
\begin{figure}[ht]
\begin{center}
\includegraphics[width=0.8\textwidth]{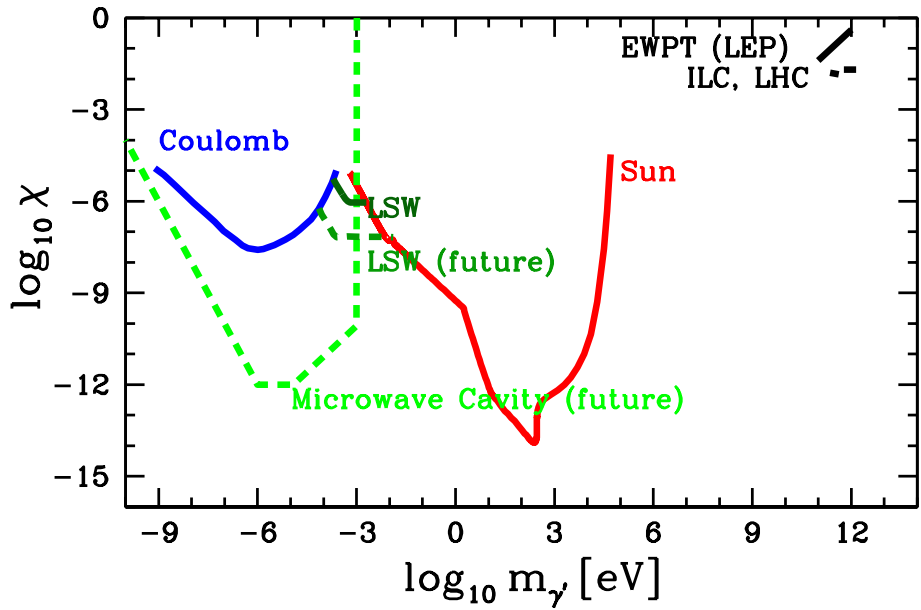}
\caption{Limits on the kinetic mixing parameter ($\chi$), corresponding to $\epsilon$ in the text, versus 
dark photon mass $m_{\gamma^\prime}$.  The figure is from  Ref.~\cite{Abel:2008ai}.
\label{epslim}}
\end{center}
\end{figure}
\begin{figure}[ht]
\begin{center}
\includegraphics[width=0.7\textwidth]{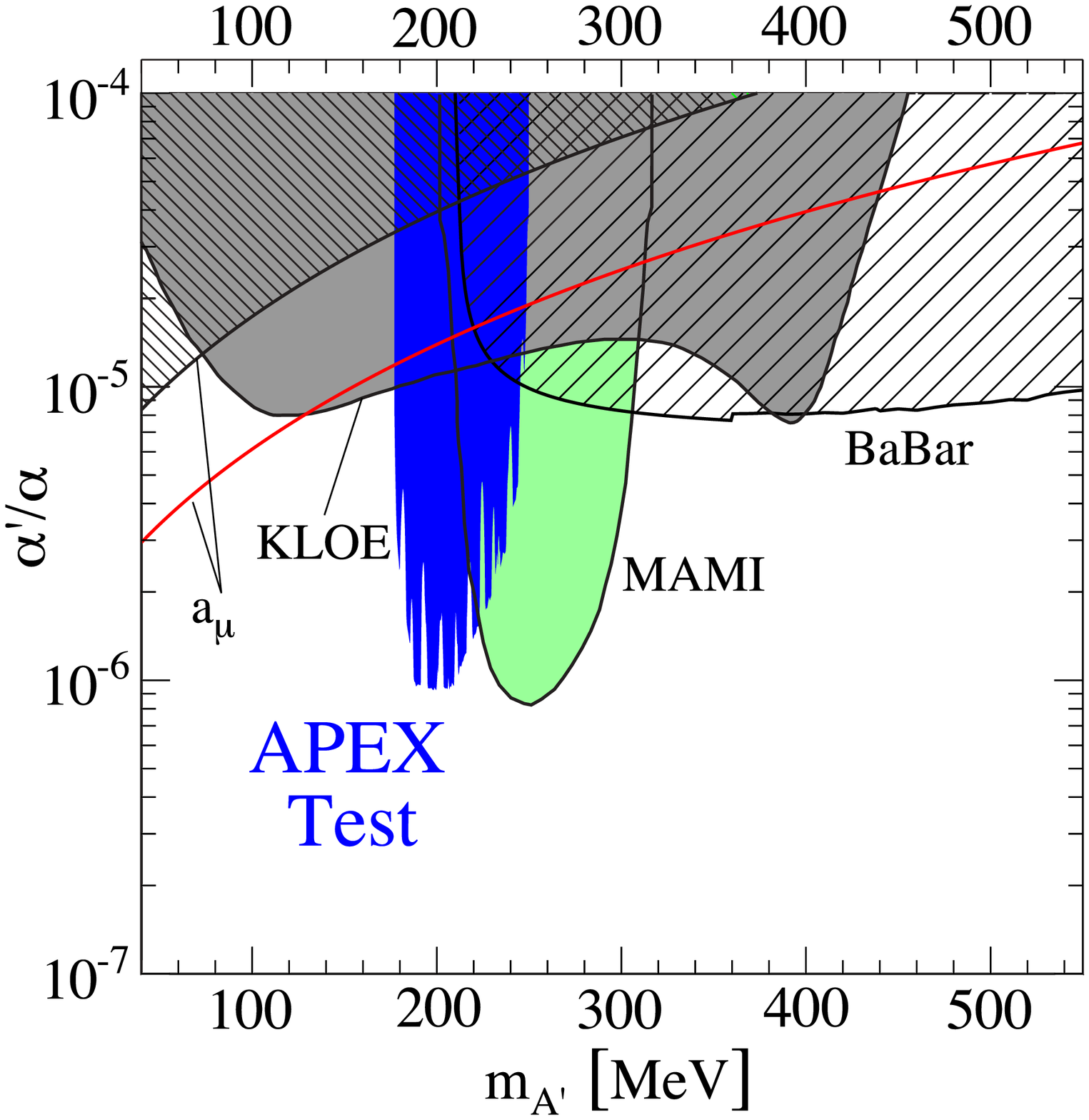}
\caption{The $90\%$ confidence level upper limits on the kinetic mixing parameter, 
where $\alpha'/\alpha=\epsilon^2$, versus 
dark photon mass $m_{\gamma^\prime}$.  The most recent results from the APEX \cite{Abrahamyan:2011gv} (shaded blue) 
and MAMI \cite{Merkel:2011ze} (shaded green) experiments are shown.  The figure is from 
Ref.~\cite{Abrahamyan:2011gv}. 
\label{apexfig}}
\end{center}
\end{figure}

There are also laboratory limits from  a generation-regeneration experiment using the ``light shining through a wall (LSW)" technique in which regenerated photons are searched for \cite{Afanasev:2008fv}. The basic idea here is that if light transforms via its mixing to a dark photon, it will not interact (or very weakly interact) with matter and can therefore pass through a ``wall" and be visible once it reappears on the other side of the wall. Such experiments \cite{Pugnat:2007nu,Ehret:2010ki}
lead to an upper limit of  $\epsilon \leq 10^{-7}$ for $m_{\gamma^\prime}$ between $10^{-5}$ to $10^{-2}$ eV. Other astrophysical as well as laboratory limits are 
summarized in Fig. \ref{epslim} (taken from Ref.~\cite{Abel:2008ai}). 

During the past few years, motivated by the interest from the dark matter related ideas \cite{Bjorken:2009mm}, searches for the photon-dark photon mixing in accelerator experiments  have been conducted \cite{Merkel:2011ze,Giovannella:2011nh,Abrahamyan:2011gv} 
and new limits have been obtained; see Fig.~\ref{apexfig} (from Ref.~\cite{Abrahamyan:2011gv}). 
The idea here is to conduct electron scattering and look for
$e^+e^-$ in the final state with different invariant masses corresponding to the dark photon mass. Since the mixing parameter  is small, the cross section for coherent electromagnetic production of the $\gamma'$ boson can be enhanced by a factor $Z^2$ by choosing a heavy nucleus as the target . The subsequent decay of the $\gamma^\prime$ boson to a lepton pair is the signature of the reaction.

\subsubsection{Massless dark photon and mini-charged matter}

If the dark photon has kinetic mixing with the familiar photon and is 
massless, dark matter acquires a small amount of familiar electric charge and can therefore have
interactions with familiar matter. The amount of charge in the dark matter (called mini-charge below)  
is proportional to  the 
photon-dark photon mixing parameter $\epsilon$. The mini-charged dark particles, in a certain mass range, can 
manifest themselves in many astrophysical settings, {\it e.g.} supernova explosions,  
as well as laboratory experiments, leading to  constraints on 
$\epsilon$ \cite{Mohapatra:1991as}.  Various experimental data  
yield $\epsilon \lsim 10^{-5}$ for mini-charged particles of mass at or below 1~eV, and $\epsilon \lsim 10^{-6}$ 
for much smaller masses; for a recent review and more details see Ref.~\cite{Jaeckel:2007hi}.
Such mini-charged particles could have implications for supernova observation if 
they have very low masses (less than a few MeV) and can be emitted 
during the supernova explosion.  This could affect the supernova 
luminosity for which there exist good estimates from the neutrino observations in SN1987A. 
These considerations put upper limits on the magnitude of the 
mini-charge {\it i.e.} the range $10^{-9} \leq Q\leq 10^{-7}$ 
is excluded for masses less than 10-20 MeV\cite{Mohapatra:1990vq}.  If the mini-charge value 
is $\geq 10^{-7}$, then minicharges  get trapped in the neutrino 
sphere and thermalize. As a result, they do not get out of the supernova in large amounts 
and the luminosity constraint is avoided.


\subsection{Collider Searches}

In many models for ADM, dark matter may either share some of the SM quantum numbers or it may interact with particles that are SM active.  Depending on the embedding of the mechanism, one can expect a number of generic signals.  For example, 
in supersymmetric contexts it is generally expected that various super-partners will emerge at  the weak scale.  Similarly, in models based on technicolor we may expect to find techni-hadrons at the TeV scale.  However, there are also specific signals 
that arise in some models.  For example, in Ref.~\cite{Kaplan:1991ah} 
the possibility of a fourth family was considered that would 
include $(t',b')$ quarks.  In Ref.~\cite{KL1} a new color-charged particle emerges that could lead to 
signals like those from long-lived or stable gluinos in supersymmetric scenarios.  Other examples include 
exotic color-charged scalars in Ref.~\cite{gu}, or exotic quarks of Ref.~\cite{dutta} may be pair produced 
through QCD interactions and decay into jets and DM particles, {\it i.e.} missing energy.  In models 
where ``neutron operators" of the type $u^c d^c  d^c$ couple directly to the hidden sector one may expect 
mono-jet plus missing energy signals at the LHC, depending on the strength of such couplings \cite{IND}.  When the 
up-type quark is a top quark, it is possible to have interesting 
mono-top plus missing energy signals \cite{IND,Andrea:2011ws,Kamenik:2011nb} that could be accessible at the LHC.
Generally speaking, there may also be particles in the theory to which the SM Higgs boson can decay, 
such as invisible states or unstable lighter scalars.

\subsection{Novel Probes}

An important aspect of ADM models is that they could motivate new ways of looking for DM that may not have been 
considered before.  An interesting example of a possible new search avenue is provided by the model in Ref.~\cite{hylo}, 
where there is no violation of generalized (dark plus visible) baryon number, yet exchange of baryon number between 
ADM and visible baryons is allowed, albeit with a small rate.  In particular, dark matter could scatter from ordinary matter 
and lead to destruction of ordinary nucleons, thereby transferring  baryon number into the dark sector.  This process was dubbed 
induced nucleon decay (IND) in Ref.~\cite{hylo}.   IND processes yield an effective lifetime $\tau_{\rm eff}$ for nucleons, depending 
on the DM density at the position of the nucleon, and can lead to signals in nucleon decay experiments 
\cite{Shelton:2010ta, hylo,IND}.  

In Refs.~\cite{hylo,IND}, using chiral perturbation theory methods, 
it was estimated that the effective life-time of a nucleon on Earth, with a local DM density of 
$\rho_{\rm DM}= 0.3$~GeV/cm$^{-3}$, is 
$\tau_{\rm eff}\sim 10^{32}$~yr, if the mass scale suppressing the dim-7 
baryon-number-transfer operator is $\sim 1$~TeV.  Such values of $\tau_{\rm eff}$ are indeed 
close to the current bounds from experiments, like  
Super-Kamiokande \cite{Kobayashi:2005pe}, suggesting that current or future nucleon decay experiments  
may be inetersting probes of certain ADM models.  In these models, 
the IND final state includes a meson and an anti-dark matter particle, mimicking standard nucleon 
decays into a meson and a neutrino.  Note however that bounds from nucleon decay experiments 
do not directly apply to IND processes, given that the kinematics of 
the two processes could be quite different.    Standard nucleon decays are typically characterized by 
meson momenta of order 300-400~MeV, while the models in Refs.~\cite{hylo,IND} 
the outgoing IND meson has a momentum $p_M \sim 600-1400$~MeV, depending on whether the process is an up-scattering 
into a a heavier dark state, or a down-scattering into a lighter state.  Such differences in kinematics 
are useful in distinguishing IND events from standard nucleon decays, but could also affect efficiency 
of event identification, due to the larger boost of the meson resulting in the collimation of its decay products 
or extra \v{C}erenkov radiation \cite{IND}.  Definitive bounds on these models then likely require a reanalysis of the available 
data.  We also note that chiral perturbation methods are only expected to yield reasonable order-of-magnitude 
estimates for IND rates, since the momenta of the mesons are $\sim 1$~GeV and an expansion  
in $p_M/\Lambda_{\rm had}$, where $\Lambda_{\rm had}$ is a hadronic scale of order 1~GeV, is not very reliable.

An indirect probe of mirror dark models based on leptogenesis  as discussed above 
(and in fact any asymmetric dark matter that uses $B-L$ changing interactions)
is via searches for neutron-anti-neutron ($N-\bar{N}$) oscillations \cite{nnbar} in reactors. If 
$N-\bar{N}$ oscillations are observed at currently accessible sensitivities, this would mean that these $\Delta B = 2$
transitions would have large enough strength to be in equilibrium in the early universe till below the electroweak phase transitions.  This  in combinations with sphalerons will erase all
preexisting baryon asymmetry in the universe and a new mechanism to generate baryons below the
sphaleron decoupling must be invoked; see for example Ref.~\cite{Babu:2006xc}. 
One would then lose the connection between the dark matter density to baryon density.
Observation of $N-\bar{N}$ oscillation will therefore rule out this scenario.  Thus a search for $N-\bar{N}$ 
oscillation could provide some essential information on the origin of dark matter.
On the other hand,  since leptogenesis is at the core of this idea,
some way to support leptogenesis is essential for this mechanism to be viable.

\section{Astrophysical Implications of Asymmetric Dark Matter}
\label{astro}

Dark matter in the present Universe is most likely to collect inside 
massive astrophysical bodies such as stars, neutron stars {\it etc.} due to its gravitational interactions as well as scattering 
on the baryons inside them. In contrast with the standard supersymmetric WIMP dark matter for which dark matter pairs annihilate to leptons and neutrinos, the asymmetric dark matter will collect 
inside the stars over the lifetime of the Universe, as a result of gravitational capture. The presence of DM 
in significant amounts could affect the properties of stellar objects.  
Such effects have been considered in several 
papers \cite{Griest:1986yu,McDermott:2011jp,Frandsen:2010yj,IND,Zentner:2011wx} 
following the classic work of Spergel and Press \cite{Spergel:1984re}. 
It has been shown that this could affect the transport properties in 
the interior of the Sun and possibly resolve the  composition anomaly 
\cite{Frandsen:2010yj} which poses conflict between the
helioseismological observations and solar composition \cite{PenaGaray:2008qe}.  

In Ref.~\cite{IND}, the destruction of stellar baryons, via IND processes, 
in models of the type proposed in Ref.~\cite{hylo} was studied.  Such processes provide extra sources of stellar heating.  However, generally speaking, effects of stellar baryon destruction were typically 
found to be negligible unless the density of the DM at the location of the star is extremely large, 
$\rho_{\rm DM}\gsim 10^{10}$~GeV/cm$^{-3}$.  Potential bounds form white dwarf heating could be an 
interesting probe of such models, but subject to uncertainties in the value of $\rho_{\rm DM}$ at the location of 
the stellar object.

Another possible effect could occur in neutron stars if the asymmetric dark matter is a scalar particle with low mass (in the 5-15 GeV range). In this case enhanced Bose condensation of gravitationally captured dark matter in neutron stars could speed up the formation of Black holes \cite{McDermott:2011jp} if the dark matter neutron cross section is larger than $10^{-47}$ cm$^2$. This will lead to a reduction in the population of neutron stars. 
For other effects of ADM on the mass and size of neutron stars, 
see Ref.~\cite{teplitz}. In particular, if the dark matter mass is lower 
than that of baryons, the ground state of the mixed dark matter neutron 
star could be higher than the Chandrasekhar limit. This would then predict that
neutron star masses higher than Chandrasekhar mass should exist in nature.


\section{Summary and Concluding Remarks}
\label{summary}

In this article, we have provided an overview of particle physics models in which dark matter abundance is set by an asymmetry 
that is related to that of baryons in the visible sector.  Such proposals provide an interesting resolution of the puzzle as to why baryon and dark matter energy densities are of similar magnitudes.   Asymmetric dark matter (ADM) models can lead to novel effects that may provide new avenues for their detection, as we have discussed here.  These models can also have interesting astrophysical implications since as they accumulate in stars, they do not self-annihilate and may lead to altered stellar dynamics.  Depending on the detailed nature of the ADM, it may affect the size and luminosity properties of neutron stars.  In some cases, ADM can annihilate ordinary baryons, which leads to heating 
of stellar objects via baryon destruction.   In some ADM models, a number of 
interesting and testable predictions emerge for collider physics, 
providing a complementary handle on these proposals.

ADM models will become compelling if it turns out that evidence builds in favor of light (5-10~GeV) dark matter and other observations put models with light symmetric (WIMP-like) candidates under stress, as has been argued in Ref.~\cite{Lin:2011gj}. Note that in the context of MSSM, a light WIMP dark matter (with mass less than about 20 GeV) is already not favored \cite{Hooper:2002nq}.  In any event, the nature of dark matter remains unknown and, in the absence of any clear 
experimental signal, examination of new scenarios that motivate alternate search 
strategies is well-worth the effort.  In this regard, ADM models, a sample of which we discussed in this brief review, 
deserve attention and can lead to a more comprehensive theoretical,  as well as experimental, approach 
to the mystery of dark matter.

\section{Acknowledgments} 
We thank Hai-Bo Yu for discussions.  The work of H. D. is supported by the United States Department of Energy
under Grant Contract DE-AC02-98CH10886.  The work of  R. N. M. is supported  
by National Science Foundation grant number PHY-0968854.


\section*{References}

\end{document}